# Electronic structure and properties of (TiZrNbCu)$_{1-x}$Ni$_x$ high entropy amorphous alloys

Katica Biljaković[a], György Remenyi[b], Ignacio A. Figueroa[c], Ramir Ristić[d], Damir Pajić[e], Ahmed Kuršumović[f], Damir Starešinić[a], Krešo Zadro[e], Emil Babić[e]

[a] Institute of Physics, Bijenička cesta 46, P. O. Box 304, HR-10001 Zagreb, Croatia

[b] Institut Neel, Universite Grenoble Alpes, F-38042 Grenoble, France

[c] Institute for Materials Research-UNAM, Ciudad Universitaria Coyoacan, C.P. 04510 Mexico D.F., Mexico

[d] Department of Physics, University of Osijek, Trg Ljudevita Gaja 6, HR-3100 Osijek, Croatia

[e] Department of Physics, Faculty of Science, Bijenička cesta 32, HR-10002 Zagreb, Croatia

[f] Department of Materials Science & Metallurgy, University of Cambridge, 27 Charles Babbage Road, Cambridge CB3 0FS, UK

Corresponding author: Ramir Ristić ; e-mail: ramir.ristic@fizika.unios.hr

Abstract

A comprehensive study of selected properties of four (TiZrNbCu)$_{1-x}$Ni$_x$ (x ≤ 0.25) amorphous high entropy alloys (a-HEA) has been performed. The samples were ribbons about 20 μm thick and their fully amorphous state was verified by X-ray diffraction and thermal analysis. The surface morphology, precise composition and the distribution of components were studied with a Scanning electron microscope (SEM) with an energy dispersive spectroscopy (EDS) attachment. The properties selected were the melting temperature ($T_m$), the low temperature specific heat (LTSH), the magnetic susceptibility $\chi_{exp}$ and the Young's modulus (E). Whereas LTSH and $\chi_{exp}$ were measured for the as-cast samples, E was measured both for as-cast samples and relaxed samples (after a short anneal close to the glass transition temperature). The LTSH showed that the electronic density of states at the Fermi level, $N_0(E_F)$, decreases with increasing x, whereas the Debye temperature ($\theta_D$) increases with x. This is similar to what is observed in binary and ternary amorphous alloys of early transition metals (TE) with late transition metals (TL) and indicates that $N_0(E_F)$ is dominated by the d-electrons of the TE. The LTSH also showed the absence of superconductivity down to 1.8K and indicated the emergence of the Boson peak above 4K in all alloys. The free-electron like paramagnetic contribution to $\chi_{exp}$ also decreases with x, whereas E, like $\theta_D$, increases with x, indicating enhanced interatomic bonding on addition of Ni. The applicability of the rule of mixtures to these and other similar HEAs is briefly discussed.

1. Introduction

About ten years ago a novel alloy design based on multiple principal alloying elements in near-equiatomic ratios has been introduced [1-4]. This design abandoned a thousands of years old strategy of making alloys based on one or at most two principal components and enabled research and probable exploitation of a huge number of completely new alloys [5-6] with structures and properties which can hardly be anticipated. Thus, this strategy provides an opportunity to greatly advance our fundamental understanding of the behaviors of alloys. Accordingly, the research of these so-called high entropy alloys (HEA [3]) resulted in short time in several hundreds of research reports, several reviews of HEA literature [6-13] and a book [14]. Initially, the interest in research on HEA arose from the unusual appearance of simple solid solution phases instead of the expected mixture of intermetallic phases [3,4]. This was interpreted by Yeh and co-workers [3,7-9,14] in terms of the stabilizing effect of high configurational entropies, $\Delta S_{conf}$, in solid solution phases (hence the name HEA). In addition to the dominant effect of $\Delta S_{conf}$ for solid-solutions on the Gibb´s energy, they proposed three further core effects in HEAs: (i) severe strain of HEA lattices due to size mismatch between alloying elements, (ii) sluggish diffusion kinetics in HEA and (iii) "cocktail effects" which can result in average composite properties (i.e., the rule-of-mixtures, RoM).

Although these core effects are emphasized in most reviews of HEAs, more experimental studies of their influence are desirable [15]. In particular, there is no so far direct evidence for the dominant effect of $\Delta S_{conf}$ in formation of solid-solutions in HEAs. Indeed, although high entropy plays important role in formation and properties of HEAs, already Cantor [4] has shown that increasing the number of components from 5 (CoCrFeMnNi) to 16 and 20 in equiatomic alloys caused the transition from single phase to multiphase alloys in spite of a nearly two-fold increase in $\Delta S_{conf}$. Recent research showed quite generally that solid solution alloys become less likely as the number of alloy elements (and also $\Delta S_{conf}$) increases [5]. However, significant amount of disorder remains in HEA despite its microstructure indicating certain degree of chemical ordering, precipitation and spinodal decomposition [16]. Indeed, entropies of mixing of multiphase HEAs often tend to be higher than those of solid-solution HEAs [11].

It is clear that in any alloy the formation and stability of a given phase or phases depend primarily on inter-element interactions (e.g. [17]) which are described by the mixing enthalpy, $\Delta H_{mix}$ (or more accurately by the formation enthalpy, $\Delta H_f$ [17]). Indeed, the oldest [18] and still the most successful criteria for the discernment (and prediction, e.g. [19]) of solid-solution HEAs from multiphase ones are based on enthalpies.

A two-dimensional $\Delta H_{mix}$ (or $\Delta H_f$)-$\delta$ plot ($\delta$ is the relative atomic size mismatch [20]) is particularly instructive for classification and phase selection of HEAs. This plot shows [13,18] how HEAs evolve with increasing $\delta$ and decreasing $\Delta H_{mix}$ from single solid solutions (with small $\delta$, $\Delta H_{mix}$~0, and obeying

Vegard´s law indicative of the generalized Hume-Rothery rules [21,22]) to multiphase (intermediate) systems where solid solution(s) coexist with intermetallic compound(s) (having larger δ, and smaller $\Delta H_{mix}$ (<0)) and amorphous HEAs (with large δ and noticeably negative $\Delta H_{mix}$, in general agreement with the empirical Inoue´s rules [23] for the easy formation of bulk metallic glasses BMG). One problem with the $\Delta H_{mix}$-δ plot is significant overlap of the multiphase systems with those forming crystalline and amorphous solid solutions, which makes its predictive power quite limited. Fortunately, the overlap in the region of crystalline solid solution is smaller than that in the amorphous region, so that in this region one can even separate different lattice structures (e.g. fcc, hcp and bcc) in true single phase HEAs [11,24]. The overlap between amorphous HEAs and multiphase alloys is almost complete [13,16], thus the prediction of HE-BMGs is probably even more difficult than that in conventional binary and multicomponent metallic glasses (MG) [23,25]. This is probably one reason that, in spite of an early start [1,2], the progress in research of amorphous HEAs is much slower than that of crystalline ones (e.g. [26-32]) and that the critical thickness of HE-BMGs is generally lower than that of conventional BMGs. The enhanced thermal stability of several HE-BMGs in respect to that of similar conventional BMGs [27,28], as well as the appearance of the solid solution phases in their crystallization products [27,28,31] seem to reflect their high $\Delta S_{conf}$. Furthermore, Takeuchi et al. [32] have shown that high $\Delta S_{conf}$ stabilizes the deep eutectics and is responsible for glass formation in PdPtCuNiP alloy. However, the formation of solid solutions (in addition to intermetallics) in HE-BMGs may also reflect their equiatomic compositions which may make the formation of stable intermetallics (such as $Zr_2Cu$ compaund in Vit. 1 alloy [27]) more difficult. The enhanced thermal stability in HE-BMGs containing early transition metals [27-29] TE may result from their lower TE concentration than that in conventional BMGs based on TEs (e.g. [33] and the text below).

The conceptual understanding of both crystalline (c-) and amorphous (a-) single phase HEAs is quite limited as seen from the core effects (c-HEAs) and an excessive use of the RoM (both in c- and a-HEAs) for the calculation of their properties (e.g. [5, 12, 26, 34]). This is partially due to lack of insight into their electronic band structure (EBS) which in metallic systems determines almost all properties (e.g. [25]) of a given system. Indeed, to our knowledge there are just two measurements of the properties directly related to EBS [34, 35] and few calculations of EBS and selected properties (e.g. [36]) for c-HEAs and no such reports for a-HEAs. As a result the development of HEAs with desirable properties is still largely a trial-and-error process like the development of BMGs [25].

The importance of EBS in understanding the properties of alloys probably shows up the best in a case of MGs composed from the early and late transition metals, TE-TL. Soon after discovery of these MGs the photoemission spectroscopy (PES) revealed the split-band structure of these alloys where the electronic density of states at the Fermi level, $N_0(E_F)$, is dominated by TE d-states [37]. Thus, the effect of alloying with TL is approximately described by the dilution of a-TE [38] which simplifies the explanation of a linear variations of the majority of properties of these MGs with TL content [33]. Furthermore, $N_0(E_F)$ values of TE-rich alloys (determined from low temperature specific heat, LTSH) where higher than those of stable hcp crystalline phases of corresponding TEs [39] and were close to those calculated for hypothetical fcc structures of TEs [40]. High $N_0(E_F)$ in TE-rich MGs leads to enhanced superconductivity and magnetic susceptibility, but also to weaker interatomic bonding, thus to lower elastic modula and thermal stability. The combined studies of PES and ab-initio modelling/calculations (AIM) were also performed on Zr-base BMGs and showed that $N_0(E_F)$ is dominated by TE d-electrons (e.g. [41]). We note that a combination of PES, LTSH and AIM is the best in order to fully comprehend EBS. In particular, experimental techniques PES and LTSH reveal the variation of the density of states (DOS) with energy (PES) and the accurate value of $N_0(E_F)$ (LTSH), but cannot provide the contributions of the alloying elements to these quantities. Theory (AIM) can in principle provide all these quantities as well as the probable local atomic structure (including chemical short range order) but its results are limited by rather small size of the sample and approximations involved in a given calculation (e.g. [36, 41]).

Particularly interesting conceptually are alloy systems that depending on preparation and/or processing conditions can form either c- or a-HEA [42, 43]. These alloys offer a unique opportunity to study the impact of amorphisation (quenched-in topological disorder) on the EBS and the physical properties of HEAs. Here we present to our knowledge the first experimental study of selected properties directly related to the EBS and interatomic bonding in amorphous $(TiZrNbCu)_{1-x}Ni_x$ ($0 \leq x \leq 0.25$) HEAs. These alloys are reported to transform into a bcc solid solution (c-HEA) upon annealing around the first crystallization peak [42]. The properties studied are: the thermal stability parameters, the (LTSH), the magnetic susceptibility ($\chi_{exp}$) and the Young´s modulus (E). The LTSH showed behaviour typical for metallic glasses composed from early and late transition metals (e.g. [33]): $N_0(E_F)$ decreases with increasing x, whereas the Debye temperature $\theta_D$ increases with x. Accordingly, $\chi_{exp}$ also tends to decrease with increasing x, whereas E increases with x (enhanced interatomic bonding), but extrapolates below that of pure Ni for x=1.00. The measured properties (other than the average atomic volume and density) do not obey the RoM as is usual in TE-TL metallic glasses [33].

## 2. Experimental procedures

The ingots of five alloys in the $(TiZrNbCu)_{1-x}Ni_x$ system with x = 0, 0.125, 0.15, 0.2 and 0.25 were prepared from high purity elements ($\geq$99.8%) by arc melting (Bühler furnace MAM-1) in high purity argon in the presence of a Ti getter. The ingots were flipped and remelted five times to ensure complete melting and good mixing of components.

Ribbons with thickness of about 20 μm of each alloy were fabricated by melt spinning molten alloy on the surface of a copper roller rotating at a speed of 25 m/s in a pure He atmosphere. Casting with controlled parameters resulted in ribbons with closely similar cross-sections (~2 x 0.02 mm$^2$) and thus with the amorphous phases having a similar degree of quenched-in disorder. The as-cast ribbons were investigated by: (1) X-ray diffraction (XRD) using a computer controlled Bruker D8 Advance powder diffractometer with a CuK$_\alpha$ source, (2) scanning electron microscopy (SEM) using a JEOL JSM7600F microscope with energy dispersive spectrometry (EDS) capability (Oxford INCA X-Act), and (3) differential thermal analysis (DTA) and differential scanning calorimetry (DSC) using the TA instrument Thermal Analysis-DSC-TGA. Standard thermal measurements were performed with a ramp rate of 20K/min up to 1400$^0$C.

As a rule as-cast alloys were used for the actual measurements of the mass density, D, the LTSH, magnetic susceptibility and Young´s modulus. D was measured using the Archimedes method [33] with a estimated maximum error of ±3% due to the small mass of the very thin and narrow ribbons. LTSH measurements were performed in the temperature range 1.8-300K using a Physical Properties Measurement System (PPMS), Model 6000 from Quantum Design Inc.. The samples with mass of about 10-20 mg were in the form of discs, obtained by compressing the amorphous ribbon [44]. The magnetic susceptibility was measured with a Quantum Design SQUID based magnetometer (MPMS5) in a magnetic field B≤5.5T and temperature range 5-300K [33,45]. The measurements of Young´s modulus, E, which was calculated from the relationship E = Dv$^2$, where v is the velocity of ultrasonic waves along the ribbon, were performed both on as-cast ribbons and the same ribbons relaxed for a short time (~30s) at temperature about 10$^0$C below the glass transition temperature, T$_g$, of a given alloy. The reason for doing this is the rather strong dependence of E in amorphous alloys on the degree of quenched-in disorder (e.g. [33]). Some data relevant to our alloys are given in Table 1 and Table 2.

|         | ΔH$_{mix}$ (kJmol$^{-1}$) | δ(%) | ΔS$_{conf}$ (JmolK$^{-1}$) | VEC | Δχ    | a(k$_p$) (Å) | a$_{th}$ (Å) | D(XRD) (gcm$^{-3}$) | D(RoM) (gcm$^{-3}$) |
|---------|---------------------------|------|----------------------------|-----|-------|--------------|--------------|---------------------|---------------------|
| a-HEA1  | -35                       | 9.2  | 13.218                     | 6.5 | 0.218 | 3.24         | 3.20         | 7.04                | 7.11                |
| a-HEA2  | -40                       | 9.3  | 13.311                     | 6.6 | 0.220 | 3.22         | 3.19         | 7.10                | 7.13                |
| a-HEA3  | -49                       | 9.5  | 13.381                     | 6.8 | 0.222 | 3.20         | 3.17         | 7.16                | 7.20                |
| a-HEA4  | -55                       | 9.7  | 13.320                     | 7   | 0.224 | 3.19         | 3.14         | 7.17                | 7.28                |

Table 1 Calculated data for a-HEA1-4 alloys.

|         | γ(exp) (mJ/molK$^{-2}$) | γ(RoM) (mJ/molK$^{-2}$) | θ$_D$(exp) (K) | θ$_D$(RoM) (K) | E(exp) (GPa) | E(RoM) (GPa) | T$_m$(RoM) (K) | T$_m$(exp) (K) |
|---------|-------------------------|-------------------------|----------------|----------------|--------------|--------------|----------------|----------------|
| a-HEA1  | 4.23                    | 4.08                    | 214            | 347            | 89.6         | 116.7        | 2000           | 878            |
| a-HEA2  | 4.14                    | 4.16                    | 220            | 350            | 93.7         | 119          | 1992           | 919            |
| a-HEA3  | 4.08                    | 4.33                    | 233            | 356            | 98.2         | 123.8        | 1977           | 918            |
| a-HEA4  | 3.94                    | 4.50                    | 257            | 362            | 103.8        | 128.6        | 1961           | 938            |

Table 2 Measured vs. calculated properties for for a-HEA1-4 alloys.

## 3. Results and discussion

In Table 1 we list for all our alloys the values of the parameters: $\Delta H_{mix}$, $\delta$, $\Delta S_{conf}$, the valence electron concentration (VEC) and the electronegativity difference ($\Delta\chi$) which are commonly used for classification of HEAs. In our calculations we used standard expressions (e.g. [13]):

$$\Delta H_{mix} = \sum_{i=1, j>i}^{n} 4\Delta H_{AB}^{mix} c_i c_j \qquad (1)$$

Where $H_{AB}^{mix}$ is the enthalpy of mixing for the binary equiatomic AB alloys, n is the number of alloying elements and $c_i$ or $c_j$ is the atomic percentage for the ith or jth element,

$$\delta = \sqrt{\sum_{i=1}^{n} c_i \left(1 - r_i / \sum_{j=1}^{n} c_j r_j\right)^2} \qquad (2)$$

where $r_i$ or $r_j$ is the atomic radius for ith or jth component,

$$\Delta S_{conf} = -R \sum_{i=1}^{n} c_i ln c_i \qquad (3)$$

where R is the gas constant,

$$VEC = \sum_{i=1}^{n} c_i (VEC)_i \qquad (4)$$

where $(VEC)_i$ is the valence electron number for the ith element, and

$$\Delta\chi = \sqrt{\sum_{i=1}^{n} c_i \left(\chi_i - \sum_{j=1}^{n} c_j \chi_j\right)^2} \qquad (5)$$

where $\chi_i$ or $\chi_j$ is the Pauling electronegativity for the ith or jth element.

The values of $\Delta H_{mix}$ and $\delta$ of all our alloys (listed in Table 1) correspond to a range of intermediate/multiphase and amorphous HEAs (e.g. [13]) and the values of $\Delta H_{mix}$ agree well with those calculated previously for $(TiZrNbCu)_{1-x}Ni_x$ (A-HEA 1-4) alloys [42] with nominally the same compositions as in our a-HEA 1-4 samples. Thus, the formation of solid solutions in these alloys by using the usual thermal treatment, as suggested in [42], may not be feasible. Accordingly, we shall not discuss in any detail VEC and other, more recent parameters and criteria for the formation of a particular type of solid solution in crystalline-HEAs (e.g.[11]). We note however, that in $(TiZrNbCu)_{1-x}Ni_x$ rods [42] a bcc-HEA phase dissappeared in the alloy with x=0.25, which has the value of VEC (Table 1) beyond the threshold value for the formation of bcc phase in HEAs [13].

XRD patterns of the melt-spun $(TiZrNbCu)_{1-x}Ni_x$ ribbons shown in Fig. 1 seem to corroborate the anticipation based on the values of parameters in Table 1. The alloy with x=0 (not shown) appeared to be crystalline with dominant bcc phase (consistent with the calculated VEC=6) with the lattice parameter close to that of β-Ti and Nb and with very small amounts of tetragonal $CuTi_3$ and CuTi compounds. The other alloys with x = 0.125, 0.15, 0.2 and 0.25 (denoted thereafter as a-HEA1-4) appear to be fully amorphous with a characteristic broad first peak around $2\Theta \approx 40^0$. In what follows we discuss only the properties of a-HEA1-4. The crystallized alloys, including that with x=0 will be discussed in a separate publication. We note that the first maxima in XRD traces of a-HEAs shift with increasing x to larger values of 2Θ as could be espected due to small size of Ni atom. Further, the broad peaks in XRD traces of a-HEAs appear at similar values of 2Θ (around $40^0$ and $70^0$) as the most intense lines of the bcc phase in our alloy with x=0 and the lines of a bcc HEA phase in similar $(TiZrNbCu)_{1-x}Ni_x$ (a-HEA1-3) suction cast rods [42] with the same values of x as our a-HEA1-3 ribbons. Therefore, from the modulus of scattering vector, $k_p$, corresponding to the first maximum of the diffraction patterns in Fig. 1 we calculated the average nearest-neighbour distances, d [33,46], for all our amorphous alloys and assuming an bcc-like local atomic structure we estimated the corresponding lattice parameters, $a=2d/3^{0.5}$, and the corresponding average atomic volumes, $V=a^3/2$. Both, the values of a and V decrease approximately linearly with increasing x in a-HEA1-4 alloys and for x=0 a (Table 1) extrapolates close to that of bcc phase in alloy with x=0 (thus, close to that of β-Ti). In Table 1 we also listed the theoretical values of a ($a_{th}$) for a-HEA1-4 alloys which were calculated from the lattice parameters for the bcc phases of constituents by assuming the validity of the Vegard´s law. The agreement between the values of a and $a_{th}$ is quite good, the values of a being about 1% larger than those of $a_{th}$. This difference may arise from the less dense atomic packing in the amorphous state and /or from the errors introduced in our calculation of a from positions of the (broad) first maxima in XRD patterns.

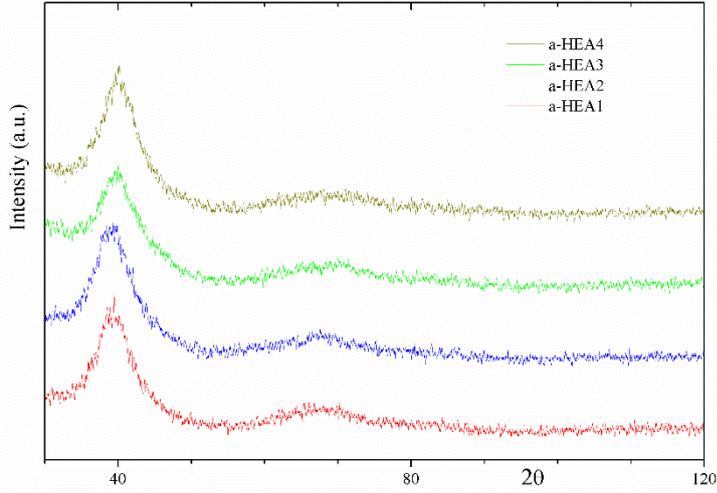

Figure 1. XRD traces of the a-HEA alloys showing amorphous halos.

Further, we used the data for average atomic volumes, V, in order to calculate the corresponding densities, D, of all our alloys [33] (Table 1). Like in the case of lattice parameters, the densities obtained from atomic volumes are about 1% lower than those deduced from the RoM (RoM, $\frac{1}{D} = \sum_i \frac{w_i}{D_i}$, where $w_i$ and $D_i$ are the weight fraction and density of the ith component, respectively), possibly indicating somewhat lower density of the amorphous phase than that of the crystalline phase. The agreement of the experimental densities with those deduced from the RoM is the consequence of the validity of Vegard´s law in our (and many other [24,46]) HEAs.The values of density obtained by using the Archimedes method also agreed quite well with the values in Table 1, but due to rather large experimental error (±3%) showed considerable scatter.Therfore in the following we use the densities determined from the X-ray results.

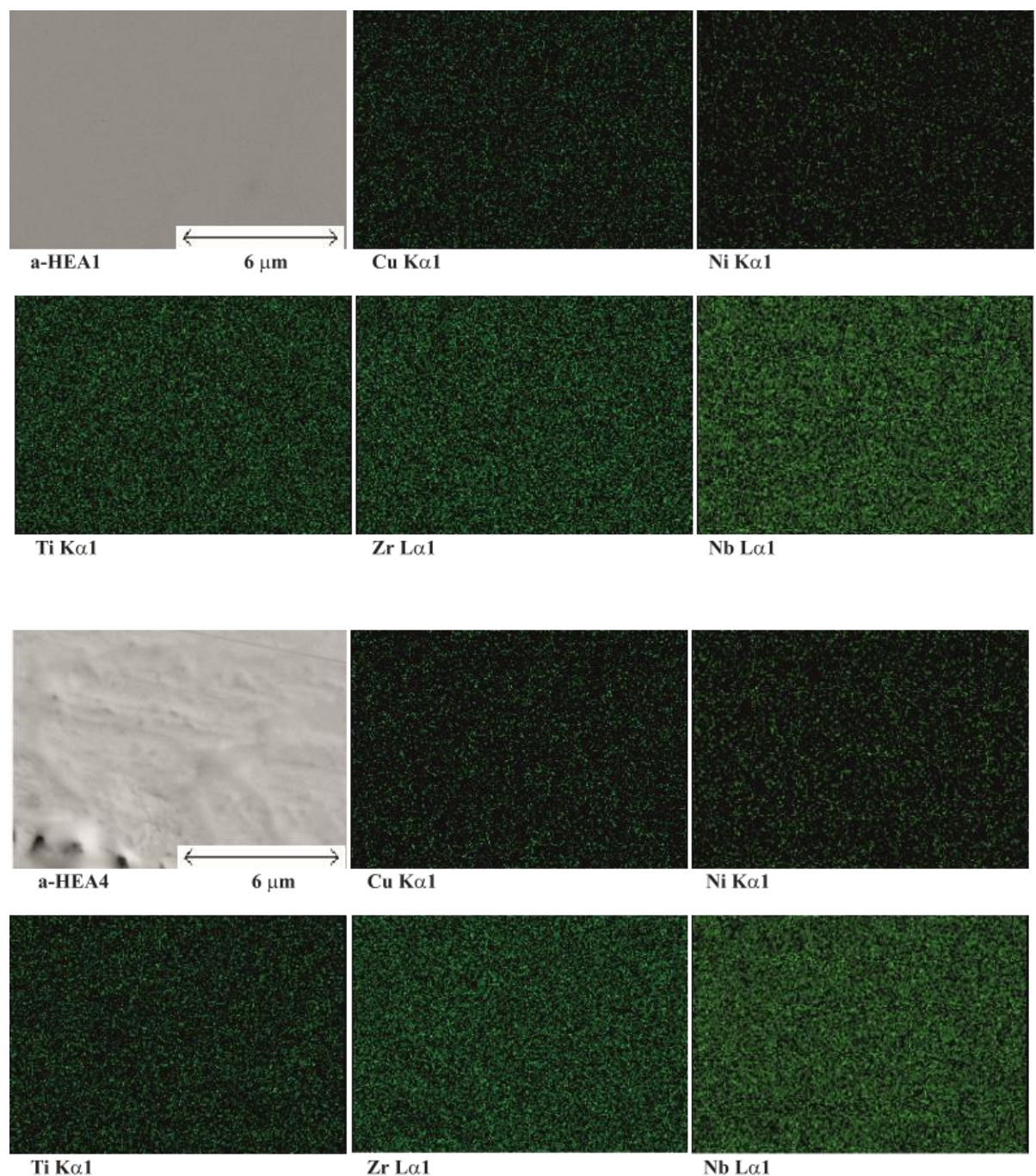

Figure 2. SEM/EDS images for a-HEA1 and a-HEA4.

Thermal analysis (TA) studies (not shown) showed well defined glass transition and crystallization start temperatures of a-HEA1-4 ribbons, $T_g$ and $T_x$ respectively, thus confirming their fully amorphous state. The values of $T_g$ and $T_x$ were similar, but about 10K higher than those reported previously for corresponding A-HEA1-4 ribbons having the same nominal compositions [42] as our alloys. In particular our DSC traces were qualitatively the same as those in Fig. 4 of Ref. 42, but our $T_g$s ranged from 390-440$^0$C, whereas $T_x$s ranged from 440-470$^0$C, both $T_g$s and $T_x$s increased with increasing x. Such influence of Ni-content on the glass formation and stability parameters in a-HEA1-4 is consistent with the corresponding variations of $H_{mix}$ and $\delta$ in Table 1, as well as with the empirical Inoue´s rules [23]. Similar variations of $T_g$ and $T_x$ with x were previously observed in binary amorphous TE-TL alloys (e.g. [33]). The novel feature of our TA studies are clear signatures of melting in all alloys, which was not reported in corresponding A-HEA1-4 ribbons [42]. The melting temperatures, $T_m$, like $T_g$ and $T_x$ increased with increasing x. The values of $T_m$ of our alloys (Table 2) are well below of that of pure copper ( which has the lowest melting point among the constituents of our alloys, $T_m$=1085$^0$C) showing strong effect of alloying on their melting. We note that at the $T_m$s the entropic contributions to the free energy are three or more times smaller than the enthalpic contributions. In Table 2 we compare the experimental $T_m$s with those calculated from the melting temperatures of constituents by using the RoM (e.g. [5]). We note that the calculated $T_m$s are about two times larger than measured $T_m$s and moreover decrease with increasing x which is opposite to the observed composition dependence. This is not surprising since the applicability of the RoM for calculation of $T_m$s of alloys is very rare. It is therefore more surprising that the RoM is almost invariably used for calculating $T_m$ in HEAs and moreover a $T_m$ obtained from the RoM is used for the selection of HEAs for specific applications (e.g. [5]). A detailed description of our TA results will be given together with the results for crystallized samples in a separate publication. Before discussing the properties of a-HEA1-4 which are more directly related to EBS and interatomic bonding it is important to address the homogeneity of our samples. Indeed, quite often the distribution of constituent elements in HEAs is uneven, and this occurs even in HEAs showing single solid solution behaviour in their X-ray patterns (e.g. [24]). Moreover, suction cast rods of A-HEA1-4 alloys [42] having nominally the same compositions as our alloys solidified to form dendritic microstructure and EDS analysis showed that the dendrites consisted mainly of Nb. The size of these dendrites was 2-5 micrometers.

In Fig. 2 we show the SEM images of a-HEA1 and a-HEA4 ribbons and the corresponding EDS mapping of the distribution of constituent elements. According to previous XRD and SEM/EDS results for as-cast rods of A-HEA alloys [42] and from our XRD result for the alloy with x=0 the elements Ni, Nb, Cu and Ti are most likely to segregate and/or to engage in the formation of intermetallic compounds in our alloys. Elemental mapping was carried out on three different places for each ribbon. The mappings were also performed on two different areas with sizes about 35x50 μm$^2$ and 7x10 μm$^2$, respectively in order to assess the eventual inhomogeneity in the distribution of constituents through the variation of composition with the size of mapped area and/or to obtain an insight into the shape and size of inhomogeneity (as was the case with dendritic segregation of Nb in A-HEA3 rod in Fig. 1 of [42]). In Fig. 2 we selected the end amorphous alloys with x=0.125 and x=0.25 in order to illustrate the evolution of the concentration of constituents with Ni-content x and also, because the low-temperature, low-field magnetization measurements of the alloy with x=0.25 revealed a small superconducting contribution with $T_c \approx 9K$ indicative of a small amount of segregated Nb. However, no such superconducting contribution was observed in other alloys and it was also absent in all samples used for LTSH measurements. As can be seen in Fig. 2 (showing the mappings on smaller areas) the distribution of selected elements was random down to a submicrometer scale in both samples. The same result was obtained for a-HEA 2 and 3 alloys and the distribution of all constituent elements was random down to submicrometer scale in all our samples . We note that statistical analysis of EDS mapping, similar to that performed in [47] would be required in order to determine more accurately the size of eventual submicrometer inhomogeneities in the distribution of elements in our samples. However, the segregation in HEAs can even occur at the nanometer scales [48,49] which cannot be probed in our experiment.

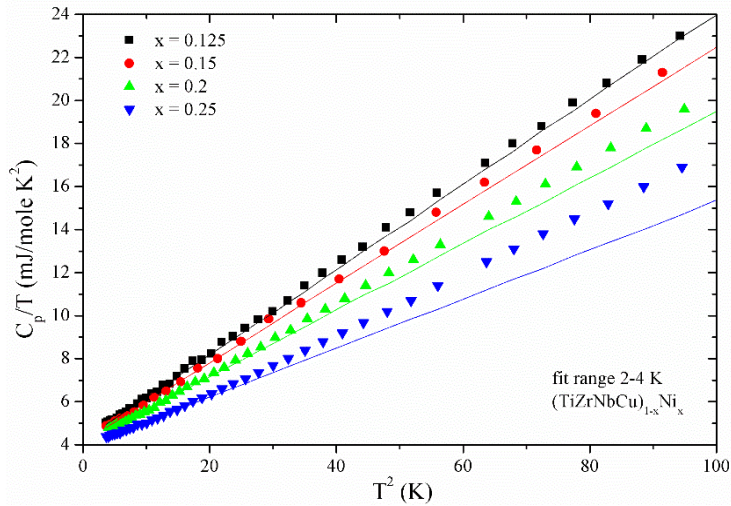

Figure 3. $C_p/T$ vs $T^2$ for $(TiZrNiCu)_{1-x} Ni_x$ alloys.

The compositions deduced from EDS at smaller areas such as those in Fig. 2 were the same to within 1 at. % as those obtained at over twenty times larger areas. The average compositions deduced from EDS agreed quite well with the nominal compositions. In particular, the concentrations of Zr and Nb , obtained from EDS analysis were practically the same as the nominal concentrations, that of Cu was 2-3 at. % larger than the nominal one and those of Ni and Ti were 1-2 at. % lower than the corresponding nominal concentrations. Considering that the probable uncertainty in concentrations determined from EDS is about ± 1 at. % we will continue to use the nominal concentrations of our alloys in further analyses.

In Fig. 3 we present the results for the LTSH of our alloys as a plot of $C_p/T$ vs. $T^2$ which is suitable for determination of both the electronic and phonon (Debye) contributions to $C_p$. These are the central results of our investigation and are to our knowledge the first such measurements performed on amorphous HEAs. The data in Fig. 3 show that up to 4K $C_p=\gamma T+\beta T^3$ and that both coefficients γ of the electronic term (intercepts in Fig. 3) and β of the Debye term (slopes of $C_p/T$ vs. $T^2$ variation) decrease with increasing x. This is the usual behaviour in amorphous alloys between the early and late transition metals (e.g. [33]) and reflects a decrease with x in the (dressed) density of electronic states at the Fermi level (Fig. 4), $N_\gamma(E_F)=3\gamma/\pi^2 k_B^2$, where $k_B$ is the Boltzmann constant, which is dominated by d-states of the early transition metals (TE) and the increase in the strength of interatomic bonding with TL content, thus an increase in the Youngs modulus and the Debye temperature $\Theta_D$ ($\beta \sim \Theta_D^{-3}$) with x. Therefore, the electronic structure and interatomic bonding in amorphous HEA composed from TE and TL seem similar to observations for binary and multicomponent TE-TL amorphous alloys, which is plausible considering the effect of dense, cubic-like atomic packing on the electronic structure of TEs [33, 39-41].

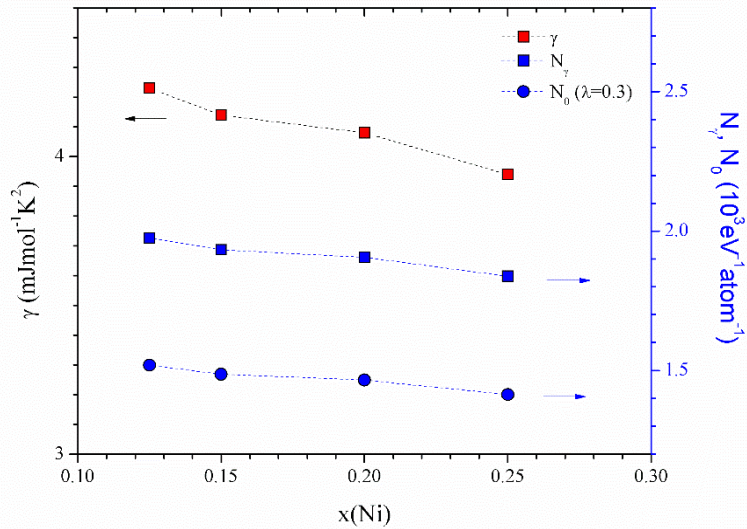

Figure 4. γ (left scale), $N_\gamma$ and $N_0$ (right scale) vs. x.

In spite of the rather high Nb content (pure Nb has $T_c$=9.2 K) and Zr and Ti components which form binary alloys with Nb with $T_c$s around 10K, the LTSH of all our alloys did not show superconductivity down to the lowest measured temperature, T=1.8K. This indicates negligible segregation of Nb in our alloys. As regards a possible formation of small disordered Nb-Zr structures, the situation is less clear since amorphous Nb-Zr films [50] showed no superconductivity down to T=1K. However, a segregation of amorphous Nb-Zr structures in our alloys seems quite unlikely considering our XRD and EDS results (Fig. 2) for these alloys and data in the literature [42]. In contrast to our experimental LTSH results which show that $T_c$<1.8K for all our alloys the values of $T_c$ for the same alloys deduced by the application of the RoM range from 4.5K to 3.9K for a-HEA-1 to a-HEA-4, respectively, which are considerably higher than the experimental results. We note that in our calculations we used consistently the values of $T_c$ for bcc phases of the superconducting components [40,50] whereas in [34] an inconsistent mixture of values of $T_c$ for hcp and bcc phases of components has been used. However, in both cases a substantial discrepancy between the experimental and the RoM values of $T_c$ has been obtained. This shows that the application of the RoM in order to deduce the values of parameters such as $T_c$ (which are closely related to the electronic band structure and lattice vibrations) of either amorphous or crystalline HEAs, yields erroneous results [34,46].

The rather high values of the density of states (DoS) of our alloys (thus high $\gamma$ and $N_\gamma$) and low $T_c<1.8$ K indicate weak electron phonon interactions, with electron phonon coupling constants $\lambda_{e-ph} \leq 0.3$ [51]. This facilates the estimation of the bare DOS at $E_F$, $N_0(E_F)=N_\gamma(E_F)/(1+\lambda_{e-ph})$ (Fig. 4) which is more directly related to the electronic band structure of an alloy than $N_\gamma(E_F)$. A similar situation, rather high $\gamma$ and $N_\gamma(E_F)$ and the practical absence of superconductivity has been observed earlier in binary Ti-TL metallic glasses [33]. In particular, in glassy Ti-Cu alloys there is no superconductivity for Cu content above 30 at. % [52] and both $\gamma$ and $N_\gamma(E_F)$ of equiatomic Ti-Cu alloy [33] are practically the same as that of our a-HEA-1 alloy. This similarity in the band structure parameters and $T_c$s of Ti-TL MGs and our a-HEA alloys is not inconsistent with the results of recent ab-initio calculations of the DOS of bcc TiZrNbMo HEA which showed the dominant contribution of 3d-electrons of Ti to the DOS at $E_F$ [36]. We note however that LTSH yields total DOS at $E_F$ only. Therefore, without the support of ab-initio calculations of DOS for our alloys we cannot be sure which constituents of our alloys give dominant contribution(s) to their $N_0(E_F)$. In Table 2 we compare experimental results for $\gamma$ of our a-HEA alloys with those calculated by using the RoM from the corresponding values for bcc phases of constituents.

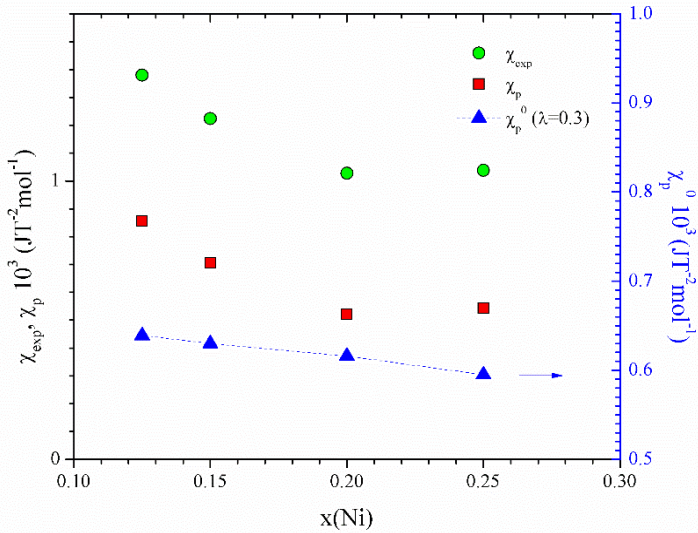

Figure 5. Magnetic susceptibilities $\chi_{exp}$ and $\chi_p$ (left scale), $\chi_p^0$ (right scale) vs. x.

We note that the calculated $\gamma_{RoM}$ are fairly close to the experimental ones which contrasts sharply with results in Ref. 34 for a similar bcc HEA, where the experimental value was much larger than the calculated one (there again an inconsistent mixture of γ values for bcc and hcp phases of the constituent elements has been used in this calculation [46]). However, the variation of the calculated $\gamma_{RoM}$ for our alloys with x is opposite to that observed, i.e. they increase with x, whereas the experimental γs decrease with x (as is usual in amorphous TE-TL alloys, e.g. [33]). Thus, the RoM seems inadequate for the calculation of the parameters related to the EBS in both a-HEAs and c-HEA [34] containing TEs. Indeed, approximating the complex effects of alloying in transition metal alloys on the EBS and superconducting parameters ($T_c$) using the RoM is reasonable only in a few special instances (e.g. [33,46]).

Before discussion of the magnetic susceptibility which is also closely related to the EBS we note that the LTSH results in Fig. 3 show that above 4K $C_p$ increases faster with temperature than the $T^3$ law predicted by the Debye model. This indicates the appearance of the so called Boson peak (BP, an excess of low-energy vibrational states with respect to that predicted by the Debye model (e.g. [53])). Thus, our results show that as in other disordered solids the BP also exists in a-HEA. Further (Fig. 3), the magnitude of the BP seems to increase with x, thus with the content of the smallest atom in our alloys. Detailed discussion of BP in our alloys will be postponed till the LTSH results for the relaxed and crystallized samples become available.

In Fig. 5 we show room temperature magnetic susceptibility, $\chi_{exp}$, of all our alloys. Like in other amorphous TE-TL alloys [33,45] $\chi_{exp}$ of our alloys showed very small (1-2%) change with temperature (5-300K) so that the room temperature values adequately represent its variation with x. Incidentally, the values of $\chi_{exp}$ in our alloys are like those for γ, in Fig. 4 similar to those of Ti-Cu MGs with Cu content less than 60 at %. Since in a-TE-TL alloys the orbital paramagnetism gives the dominant contribution to $\chi_{exp}$, this result probably indicates that the averaging of the contributions of orbital paramagnetism of transition metal components in our alloys yields value close to that of titanium. As noted earlier [25,33,45,54], in spite of apparently complex origin of the magnetic susceptibility in TE-TL MGs [55], $\chi_{exp}$ in these alloys usually decreases with decreasing TE content in qualitatively the same way as their DOS at $E_F$ (e.g. γ in [33]). This decrease reflects a linear decrease of both the orbital paramagnetism ($\chi_{orb}$) and the Pauli paramagnetism of the d-band ($\chi_p$) with increasing TL content in nonmagnetic alloys. Among these contributions only the free electron part of Pauli paramagnetism of the d-band ($\chi_p^0$) is proportional to $N_0(E_F)$. As seen in Fig. 5 $\chi_{exp}$ and $\chi_p$ ($\chi_p = \chi_{exp} - (\chi_{dia} + \chi_{orb})$) in our alloys decrease linearly with x till x=0.2, but increase a little at x=0.25. At present we have no proper explanation for this small increase of $\chi_{exp}$ at x=0.25 which may arise from the segregation of a small amount of more magnetic phase within this alloy and/or from an increase in the Stoner enhancement (e.g. [33,45]) factor (S) in this alloy. In any case this small increase of $\chi_{exp}$ in a-HEA 4 alloy is not reflecting an increase of DOS at $E_F$ as seen from the smooth variations of $N_0(E_F)$ and the corresponding $\chi_p^0$ in Figs. 4 and 5 respectively.

As noted earlier [33,54,56], there is a simple relationship between the EBS and the mechanical and thermal properties of amorphous TE-TL alloys (which is quite uncommon in crystalline metallic systems [56]). In particular, the decrease in $N_0(E_F)$ is accompanied with the increase in Young´s modulus E, Debye temperature $\theta_D$, and the thermal stability (represented by $T_g$ and $T_x$) in these alloys. Thus, the interatomic bonding increases with decreasing $N_0(E_F)$ and accordingly the stiffness and parameters related to thermal stability and the lattice vibrations increase, too. As seen from Fig. 6 our a-HEA1-4 samples conform to this pattern since their E and $\Theta_D$ (as well as $T_g$ and $T_x$) increase with x, and thus with decreasing $N_0(E_F)$ (Fig. 4). Figure 6 also shows strong effect of relaxation on E: it strongly increases in relaxed samples (as is usual in MGs, e. g.[53]) and this increase depends sensitively on the quenched-in disorder in the as-cast samples. In particular, the a-HEA2 alloy (x=0.15) showing the smallest increase of E upon relaxation was about 10% thicker (2 μm) than the other ribbons. We note that the increase of E upon relaxation in our a-HEAs ( about 10%) is about two to three times larger than that typically observed in binary TE-TL MGs [33] and also E of relaxed a-HEA samples shows simple, nearly linear variation with x. A linear extrapolation of this variation to x=1 yields E≈185 GPa for amorphous Ni which is well below that for crystalline Ni (E≈200 GPa). Moreover, as seen in Table 2 the experimental values of E generally do not agree with those calculated by using the RoM. In particular, the calculated values are 25-30% larger than the measured values for relaxed samples and also show weaker variation with x than the experimental results. Therefore, in contrast to the claims in literature the RoM is not suitable for a calculation of the mechanical properties in either amorphous [26] or crystalline HEAs (e.g. [57]) containing TEs. Fig 6 aso shows that the Debye temperatures of our samples increase with increasinng x as does E. The total magnitude of this increase is about 20%, similar to that of E. Since the decrease in the average atomic mass of our alloys with x is small (≤2.5%) a strong increase of $\theta_D$ with x is mostly due to the enhancement of the interatomic bonding. As seen from Table 2 the RoM is not adequate for calculating $\theta_D$ of our alloys. In particular, in our alloys as in bcc HEA from Ref. 34 the RoM yields much higher Debye temperatures than the measurements, and also the relative increase of the calculated temperatures with x is much smaller (only about 4%) than the observed one. Thus, the RoM gives poor description of the vibrational properties of both amorphous and crystalline [34] HEAs.

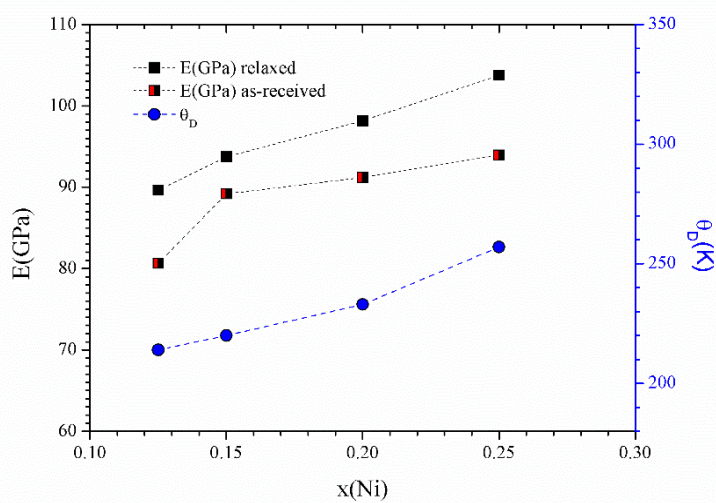

Figure 6. Youngs modulus (left scale) and $\theta_D$ (right scale) vs. x .

4. Conclusion

A comprehensive study of the electronic structure (EBS) and selected properties of $(TiZrNbCu)_{1-x}Ni_x$ (x<0.25) amorphous high entropy alloys (a-HEA) has been performed. The properties selected were the density of electronic states (DoS) at the Fermi level ($E_F$), thermal stability parameters ($T_g$, $T_x$ and $T_m$ :the glass transition, crystallisation and melting temperatures, respectively) as well as vibrational ($\theta_D$, the Debye temperature), magnetic ($\chi_{exp}$, the magnetic susceptibility) and mechanical/elastic properties (E, Young´s modulus). Such studies are important since they can reveal the relationship beetwen the EBS (in metallic systems almost all properties derive from EBS) and other properties of these complex new materials. In spite of their importance, such studies are so far rare for crystalline HEAs [34,35] and absent for a-HEA. Therefore, we hope that our study may stimulate future studies of the EBS-property relationship in HEAs which can provide a better understanding of these systems than that obtained from the often unsuccessful application of the rule of mixtures (RoM) and from the postulated four core effects (e.g. [11,12,14,26]).

The main result of our study (obtained from low temperature specific heat measurements, LTSH) is that in all HEAs studied the electronic density of states at the Fermi level, $N_0(E_F)$, is probably dominated by the d-electron states of the early transition metals (TE). Since the same phenomenon occurs in the binary and multicomponent metallic glasses (MG) based on TEs [25,41], the properties of our a-HEAs behave in qualitatively the same way as those in TE base MGs (e.g.[33]). In particular, $N_0(E_F)$ (thus also $\gamma$ and $\chi_p^0$, the coefficient of the linear term in LTSH and the free electron like Pauli paramagnetic contribution to magnetic susceptibility, respectively) decrease with increasing content of late transition or noble metals (x), whereas the properties related to interatomic bonding (e.g. $T_g$, $T_x$, $T_m$, E and $\theta_D$) increase with x. Our LTSH measurements also provided the first evidence that the Boson peak is present in a-HEAs, too.

Apparently, under the circumstances met in our a-HEAs (and probably all other HEAs containing significant fraction of TEs) where the DoS at and around $E_F$ is probably dominated by the TE component(s) only, the applicability of the rule of mixtures (RoM) (which gives equal weight to the contributions of all components) in modelling their EBS and other parameters and properties is highly unlikely. Indeed, in our a-HEAs, like in TE-containing MGs (e.g.[33]) the RoM gives inaccurate predictions for all their properties other than their atomic volumes and densities. (The variation of average atomic volumes according to Vegard´s law leads automatically to the mass density obeying the RoM.). In particular, in the present alloys the RoM predicts erroneous variation of the DoS at $E_F$ with Ni content x (an increase, rather than the observed decrease of $N_0(E_F)$ with x) and largely overestimates their superconducting trasition temperatures, $T_m$ and E. We believe that, as in the case of a bcc $Ta_{34}Nb_{33}Hf_8Zr_{14}Ti_{11}$ HEA [34] the RoM will be inadequate for understanding the EBS and other properties of crystalline TE-containing HEAs with cubic crystalline phases, too. Thus, the present extensive use of the RoM for the calculation of various properties of HEAs (other than atomic volumes and densities for single phase alloys), as well as the use of the RoM for the selection of HEAs for particular applications should be reconsidered (at least in the case of TE-containing HEAs).

Acknowledgement

We thank Prof. J.R. Cooper for useful comments and Dr. O. Lozada for the help with DSC/TGA.